# Feature Homomorphism - A Cryptographic Scheme For Data Verification Under Ciphertext-Only Conditions

Huang Neng
FuZhou, FuJian, China
hn1014@163.com

**Abstract**: Privacy computing involves the extensive exchange and processing of encrypted data. For the parties involved in these interactions, how to determine the consistency of exchanged data without accessing the original data, ensuring tamper resistance, non-repudiation, quality traceability, indexing, and retrieval during the use of encrypted data, which is a key topic of achieving "Data Availability versus Visibility". This paper proposes a new type of homomorphism: Feature Homomorphism, and based on this feature, introduces a cryptographic scheme for data verification under ciphertext-only conditions. The proposed scheme involves designing a group of algorithms that meet the requirements outlined in this paper, including encryption/decryption algorithms and Feature Homomorphic Algorithm. This group of algorithms not only allows for the encryption and decryption of data but also ensures that the plaintext and its corresponding ciphertext, encrypted using the specified encryption algorithm, satisfy the following property: the eigenvalue of the plaintext obtained using the Feature Homomorphic Algorithm is equal to the eigenvalue of the ciphertext obtained using the same algorithm. With this group of algorithms, it is possible to verify data consistency directly by comparing the eigenvalues of the plaintext and ciphertext without accessing the original data (i.e., under ciphertext-only conditions). This can be used for tamper resistance, non-repudiation, and quality traceability. Additionally, the eigenvalue can serve as a ciphertext index, enabling searchable encryption. This scheme completes a piece of the puzzle in homomorphic encryption.

**Keywords**: Privacy Computing, Data Consistency, Searchable Encryption, Zero-Knowledge Proof, Feature Homomorphism

## 1. Introduction

Nowadays, with the rapid development of Internet and mobile Internet technology, information security and privacy protection issues are receiving increasing attention. Countries around the world are enacting laws to protect information security, with the goal of mining data value, serving customers, and promoting economic development while safeguarding information security and privacy. To achieve the goal of "Data Availability versus Visibility" privacy computing has become one of the hottest research topics. During the process of privacy computing, encrypted data is repeatedly exchanged, and for all parties involved, it is also an important topic in the digital economy to verify Data Consistency, Searchable Encryption, and Zero-Knowledge Proof without obtaining the original data.

Current data consistency verification schemes mainly use hash algorithms. By calculating the hash value of the data and comparing it with the original hash value provided by the data provider, if the hash value matches the one given by the data provider, it can be judged that the data is consistent with the data provided by the provider and there is no loss or tampering of the data. One of the necessary conditions is that the consistency verifier must obtain the original data (or plaintext). As shown in Fig 1 (the process of consistency verification is shown within the red frame, which requires decryption of the ciphertext).

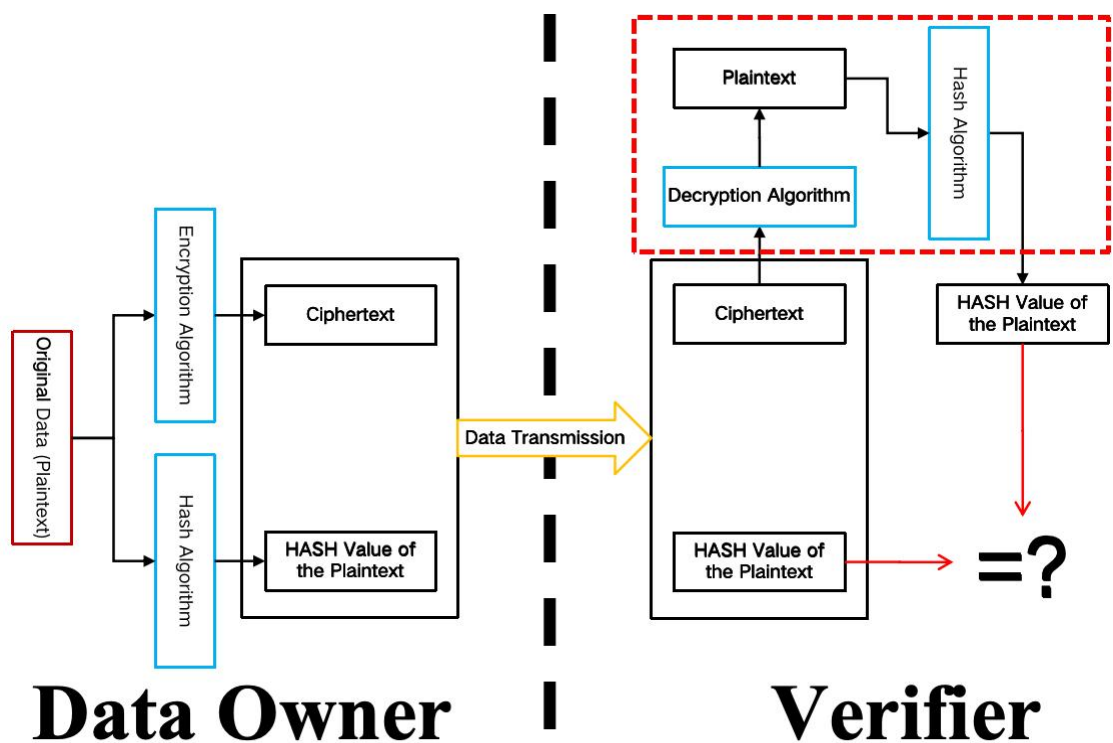

Fig 1 Traditional Data Consistency Verification Diagram

Additionally, there are schemes that use privacy computing related technologies, in the form of Zero-Knowledge Proofs, to indirectly prove data integrity and consistency.

This paper proposes a group of algorithms for data verification under ciphertext-only conditions, without decryption, to address issues such as verify data consistency without accessing the original data (plaintext). The algorithm aims to solve the following problems:

First, it is necessary to achieve the basic functions of an encryption algorithm, namely the ability to encrypt data (plaintext) and successfully decrypt the encrypted ciphertext back to the original plaintext. This is a fundamental capability that any general encryption and decryption algorithm should possess.

Second, it should be able to determine that the original data corresponds to and is consistent with the encrypted data without accessing the original data (i.e., that the encrypted data was indeed encrypted from the original data). This means being able to verify plaintext consistency without decrypting the ciphertext (i.e., without accessing the original data).

To achieve the data consistency verification under ciphertext-only conditions as proposed in the second point above, this paper proposes to constructing encryption and decryption algorithms through transformations between mathematical structures in geometry, algebra, and number theory. Then, the invariants in these tmathematical transformations are used as verification codes for plaintext-ciphertext data consistency. Here, we define the relevant concepts of this scheme:

**Feature Homomorphism**: If a mapping from plaintext space to ciphertext space ensures that the one-to-one mapped plaintext and ciphertext pairs simultaneously satisfy a certain property, and this property allows consistent eigenvalues to be calculated from both plaintext and ciphertext, then the mapping is said to satisfy Feature Homomorphism, or to possess Feature Homomorphism.

The cryptographic method that meets this scheme is called "Feature Homomorphic Encryption", the group of algorithms is called the "Feature Homomorphic Algorithm Group", and the encryption and decryption algorithms with feature homomorphism are named "Feature Homomorphic Encryption Algorithm" "Feature Homomorphic Decryption Algorithm", and "Feature Homomorphic Algorithm (FHA)" respectively. The eigenvalue is referred to as the "Homomorphic Feature Value (HFV)".

This scheme can also be used for tamper resistance, non-repudiation, quality traceability, and as an index for ciphertext, enabling ciphertext retrieval. Additionally, it can be applied in the field of zero-knowledge proofs, falling within the scope of Privacy Computing.

This scheme is related to two fields. The first is homomorphic encryption, which allow for certain operations (such as addition, subtraction, multiplication, or division in a broad sense) to be performed on encrypted data, the output of these operations, when decrypted, yields the same result as if the operations were performed on the original, unencrypted data. The similarity with this scheme lies in the use of certain mathematical structures (algebra, geometry, number theory) to construct algorithms that preserve some properties during structural transformations (or mappings). Additionally, this method can serve as a way to implement searchable encryption.

The second is zero-knowledge proof, where the prover convinces the verifier that a certain statement is true without providing any useful information to the verifier. Therefore, this scheme essentially proves to the verifier, without accessing the original plaintext data, that the ciphertext is indeed encrypted from the original plaintext data. In fact, this is a form of zero-knowledge proof.

## 2. Related Work

As mentioned in the previous chapter, this scheme is a branch of privacy computing. Current research in privacy computing can be categorized into the following types.

## 2.1. Technologies And Tools

Technologies and tools for implementing privacy computing. First involve Data Anonymization, ensuring that privacy is not compromised under various conditions and scenarios. These technologies include Data Obfuscation, Steganography, and Anonymous Communication. Evans et al[1] discusses the technology of Oblivious Relaying, where data is transmitted through relay nodes that forward the data without decrypting its content (i.e., being oblivious to the data content). This enhances the privacy and security of data transmission. Then explores the applications of this technology in financial cryptography and data security. Narayanan et al[2] proposed a method for robust de-anonymization of large-scale sparse datasets. Based on this, the potential threats to user privacy posed by this method are discussed, and possible defense strategies and improvements are proposed to enhance the security of data anonymization and reduce the success rate of de-anonymization.Similarly, from the attacker's perspective, Melis et al[3] explores the potential inference attacks that may occur during federated learning and verifies and compares protection schemes such as Differential Privacy, Gradient Clipping, Noise Addition, and Model Compression. Chamikara et al[4] explored effective obfuscation techniques in the data mining process to protect data privacy.

Secondly, through probabilistic and statistical analysis methods, it aims to mine and obtain as much useful, effective, and real statistical information as possible from the anonymized data that contains a large amount of noise. Di et al[5] propose LDP-DL, a privacy-preserving distributed deep learning framework via local differential privacy and knowledge distillation, where each data owner learns a teacher model using its own (local) private dataset, and the data user learns a student model to mimic the output of the ensemble of the teacher models. Wood et al[6] explored how to use sampling-based inference methods for high-dimensional data publishing, aiming to achieve differential privacy protection while minimizing data loss and maintaining data utility.Dwork et al[7] introduced the concept of centralized differential privacy, using Gaussian noise or other complex probability distributions to replace traditional noise addition methods, thereby improving traditional differential privacy schemes in terms of privacy protection, noise magnitude, and data utility.

Third, conducting data analysis through cryptographic methods, with homomorphic encryption being the most typical example, allows data analysis to be performed on encrypted data, thereby avoiding privacy breaches. After years of development, international standard organizations have also established standards for homomorphic encryption[8]. Halevi et al[9] improved the RNS variant of the BFV homomorphic encryption scheme, addressing its performance and security issues. However, homomorphic encryption still faces challenges such as high resource consumption, performance, and complexity, which limit its practical application.

Fourth, other technologies that achieve specific privacy protection goals include Zero-Knowledge Proofs, Searchable Encryption, and Verifiable Computation. Garg et al[10] proposed new secure program obfuscation and functional encryption techniques using homomorphic encryption and other technologies, aim to make the program difficult to understand while maintaining its functionality and ensuring that user inputs are processed without leaking the privacy data contained within them. Zheng et al[11] proposed an encrypted keyword search scheme that can handle concurrent keyword queries, which can be applied to large-scale dynamic encrypted cloud data. Joye et al[12] proposed a privacy-preserving k-nearest neighbor (k-NN) search scheme, optimizing both data privacy and computational efficiency. Li et al[13] proposed a privacy-preserving data deduplication method in a cloud storage environment based on privacy computing.

## 2.2. Scenario-Based Solutions And Practical Applications

Further more, scenario-based implementation Solutions for privacy computing, include Secure Multi-Party Computation, Trusted Execution Environment, Federated Learning, etc., and research on how to apply the above technologies and solutions to various scenarios such as Cloud Computing, Machine Learning/Deep Learning, Big Data, and IoT. Kim et al[14] proposed a scheme and evaluation method for implementing secure logistic regression based on homomorphic encryption technology.Yang et al[15] integrated data sampling, differential privacy, data indexing, and querying technologies to propose a comprehensive tool for sub-linear privacy-preserving data analysis. Chun et al[16] introduced a local differential privacy mechanism to provide an effective federated learning method that adapts to various privacy needs while maintaining data privacy. Bonawitz et al[17] provides a secure aggregation method for privacy-preserving machine learning, particularly in federated learning scenarios, allowing multiple participants to collaboratively train a global model without sharing their individual data, thereby ensuring the privacy of model parameters during the aggregation process. Abadi et al[18] explored how to apply differential privacy techniques in deep learning to protect user data. Phong et al[19] proposed a new deep learning framework that utilizes additive homomorphic encryption to ensure data privacy throughout the entire computation process. Nasr et al[20] designed a regularizer that enables models incorporating it to resist membership inference attacks, thereby achieving membership privacy protection in machine learning. Phong et al[21] studied the privacy protection issues in the process of updating or maintaining aging generalized linear models and proposed a privacy-preserving SVM training method. Bittau et al[22] proposed the Prochlo system, which combines anonymization, differential privacy, and secure multi-party computation technologies to protect privacy throughout the entire process in large-scale population analysis scenarios. Nasr et al[23] conducted a comprehensive analysis of the privacy protection issues in deep learning, correspondingly, summarized the current countermeasures and performed analysis and validation. Gai et al[24] provided a data synchronization scheme in mobile cloud computing that combines various advanced privacy protection technologies, ensures data synchronization and consistency while effectively preventing the leakage of sensitive data. Li et al[25] explored how to utilize encryption technology, differential privacy, and secure multi-party computation to achieve privacy protection in a hybrid cloud environment while ensuring the practicality and efficiency of data mining computations. Using a similar scheme, Li et al[26] also studied the issue of privacy-preserving data integration across multiple data sources and proposed a method that combines encryption technology, differential privacy, and secure multi-party computation. Pirk et al[27] proposed a new privacy-preserving data parallel processing architecture that can efficiently process large-scale data in a public cloud while protecting data privacy. Liu et al[28] studied methods for protecting privacy in multi-party online services. Tang et al[29] proposed a privacy-preserving biometric recognition method using homomorphic encryption technology. Additionally, there are more specific applications of privacy computing in fields such as medicine, finance, and power.

## 2.3. Data Consistency And Integrity Verification Scheme

We can see that in current research, many discussions are beginning to focus on schemes for data consistency and integrity verification under privacy computing. First, apply current privacy computing technologies and tools to data consistency verification to achieve data consistency verification under privacy protection. Manulis et al[30] proposed a scheme for achieving data integrity based on fully homomorphic encryption and zero-knowledge succinct non-interactive arguments of knowledge (SNARK) technology.

Secondly, starting from distributed and multi-user application environments such as IoT, edge computing, cloustard computing, and blockchain, research data consistency and integrity. Cui et al[31] discusses methods for efficiently verifying the integrity of data in edge computing environments, It belongs to an interactive edge computing data consistency verification method. Quanyu et al[32] presents a scheme that uses blockchain technology to ensure data integrity in IoT systems while maintaining privacy, the scheme utilizes technologies including Lifted EC-ElGamal cryptosystem, bilinear pairing and aggregated signature. Youssef et al[33] using SIS problem and identity-based signatures technologies, constructed a scheme for ensuring data integrity in fog computing architectures, which is crucial for distributed data sharing. Zhao et al[34] explores data integrity verification in mobile edge computing environments, particularly in scenarios involving multiple vendors and servers. In their proposed SIA scheme, each round of verification selects unreliable data replicas based on the QoS (Quality of Service) of the caching service and the unverified time of the data replicas, in order to reduce computational and communication overhead. Tong et al[35] proposed two privacy-preserving data integrity verification schemes called ICE-basic and ICE-batch in mobile edge computing environments. These schemes are respectively used for situations where users want to check data integrity on a single edge or multiple edges. Zhao et al[36] discussed the current methods for data consistency verification in edge computing, including some of the methods mentioned earlier, and provided a comprehensive survey that examines the current research status, open problems, and future directions in the field of edge data integrity verification (EDIV) .

It can be seen that current privacy computing research focuses on how to apply it to practical scenarios. Moreover, data consistency, integrity, and verifiability under privacy computing have received attention. In response, this paper provides a cryptographic tool that can be used for verifying data consistency, integrity, and other aspects—Feature Homomorphism.

## 3. Definition of Feature Homomorphism Data Verification Scheme

This scheme requires the construction of a Feature Homomorphic Algorithm Group, including the Feature Homomorphic Encryption Algorithm (abbreviated as E), the Feature Homomorphic Decryption Algorithm (abbreviated as D), and the Feature Homomorphic Algorithm (abbreviated as FHA). The Feature Homomorphic Encryption and Decryption Algorithms implement the encryption and decryption of the original data (plaintext), while the FHA computes the FHV of both plaintext and ciphertext, achieving data verification.

### 3.1. Definition of Feature Homomorphic Algorithm Group

The encryption algorithm E, the decryption algorithm D, and the FHA involves parameters such as plaintext M, ciphertext C, encryption key $k_E$, and decryption key $k_D$. They satisfy the following relationships:

Plaintext is encrypted into ciphertext using the Feature Homomorphic Encryption Algorithm:

$$C = E(k_E, M) \tag{1}$$

Ciphertext is decrypted into plaintext using the Feature Homomorphic Decryption Algorithm:

$$M = D(k_D, C) = D(k_D, E(k_E, M)) \tag{2}$$

Plaintext and ciphertext obtain consistent HFV through the FHA:

$$HFV = FHA(M) = FHA(C) = FHA(D(k_D, C)) = FHA(E(k_E, M)) \tag{3}$$

The encryption algorithm E and decryption algorithm D satisfy the general characteristics of encryption and decryption algorithms, including:

1) **Reversibility**: The ciphertext obtained after encrypting plaintext can be decrypted back to the original plaintext.

2) **Efficiency**: The computational time complexity and space complexity of the algorithms can be controlled within a certain range, allowing encryption and decryption to be performed with limited resources within a time frame tolerable to users or business operations.

3) **Security**: The algorithms provide Computational Security, Semantic Security, and Indistinguishability.

4) **Symmetry**: The Feature Homomorphic Encryption and Decryption Algorithms can be either symmetric or asymmetric. This means that the keys and the corresponding encryption and decryption algorithms can be the same or different.

For the FHA, the following requirements must be met:

1) **Feature Homomorphism**: The computed results for both plaintext and ciphertext, i.e., the HFV, must be consistent.

2) **Irreversibility**: It should be computationally infeasible to deduce the plaintext or ciphertext from the computed results.

3) **Weak Collision Resistance**: Apart from the plaintext and ciphertext pairs obtained using the Feature Homomorphic Encryption and Decryption Algorithms, it should be computationally infeasible to find another plaintext and ciphertext pair with the same HFV.

4) **Avalanche Effect**: A slight change in the plaintext (ciphertext) data should cause a significant change in the HFV.

5) **Efficiency**: The time complexity and space complexity of the computation should be controlled within a certain range, allowing the verification algorithm to be completed within a time frame tolerable to users or business operations using limited resources.

### 3.2. Data Consistency Verification Based On FHA

The ciphertext-only data consistency verification scheme based on the FHA is shown in Fig 2:

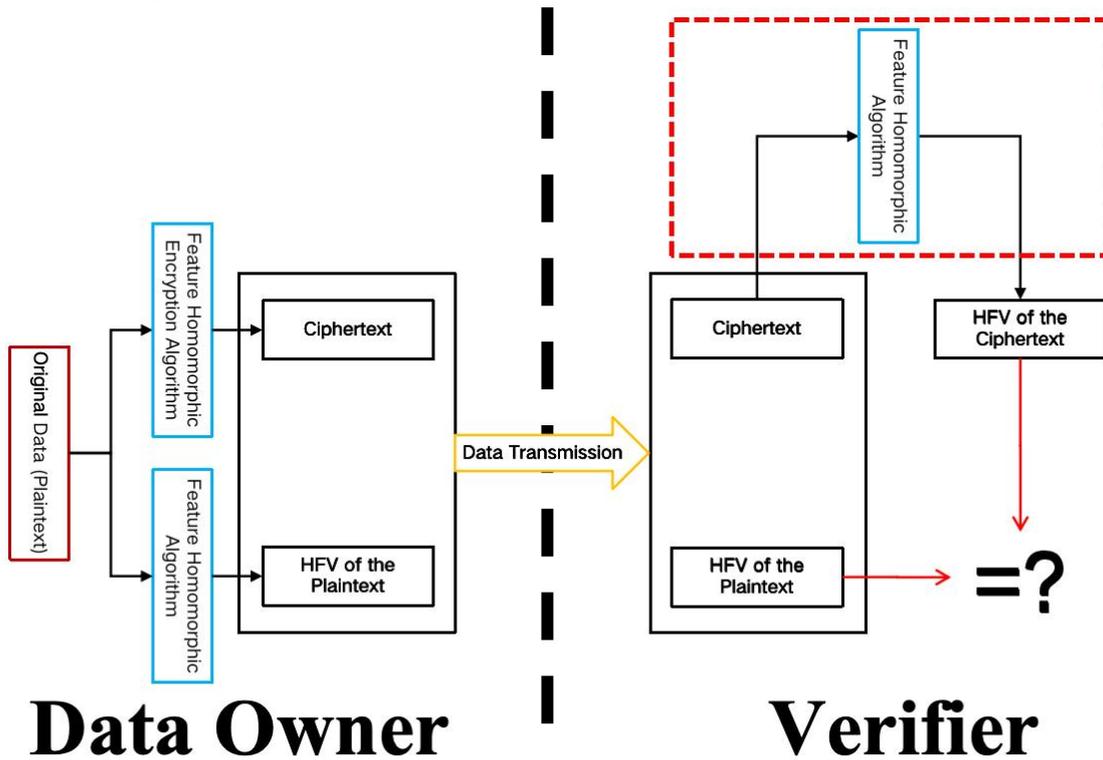

Fig 2 Schematic diagram for ciphertext-only data consistency verification scheme based on Feature Homomorphic Algorithm

Generating the HFV of plaintext through an algorithm:

$$HFV = FHA(M) \qquad (4)$$

Encrypt the plaintext and send the ciphertext along with the HFV to the verifier:

$$C = E(k_E, M) \qquad (5)$$

After receiving the ciphertext and the HFV, the verifier calculates the HFV of the ciphertext using the HFA and compares the calculated HFV of the ciphertext with the received HFV of the plaintext to check for consistency:

$$HFV = ? FHA(C) \qquad (6)$$

If they are consistent, it indicates that the ciphertext was encrypted from the plaintext with the same HFV. As seen in Fig 2, generating the HFV within the red box does not require decrypting the ciphertext.

## 4. Projective Geometry Constructs Feature Homomorphic Algorithm Group

In paper "A Comparative Review of Recent Researches in Geometry" (1872), German mathematician Felix Klein pointed out that geometry is the study of properties that remain invariant under certain transformations (also known as the "Erlangen Program"). According to this, the geometric properties of figures are equivalent to their invariance under specific transformation groups. It can be seen that geometric research precisely meets the requirements for Feature Homomorphism. As an implementation example, projective geometry is used here, which studies the geometric properties of figures that remain invariant under any projective transformation. The cross-ratio is one of the invariants under projective transformations. It is easy to see that since the cross-ratio remains invariant under projective transformations, if the constructed E and D algorithms are projective transformations, the cross-ratio can be one of the methods to construct the consistency verification algorithm FHA described in this paper. Due to the numerous types of projective transformations, using projective transformations and the cross-ratio can construct a whole cluster of E, D, and FHA algorithm groups. To illustrate the feasibility of this method more clearly, this paper first constructs the algorithm group E, D, and FHA using the simplest type of projective transformation and cross-ratio, and temporarily ignore the cryptographic properties that the algorithm needs to satisfy.

### 4.1. Projective Transformations and Cross-Ratio

The algorithm group constructed in this paper is based on projective transformations and cross-ratio. Here, we briefly introduce the relevant concepts. Projective geometry studies the positional relationships of figures, focusing on the invariant properties of figures (points, lines, etc.) when they are projected onto lines, curves, planes, or surfaces. We introduce one of the simplest projective transformations: central projection and the cross-ratio theorem. Using this central projection and cross-ratio, we construct an algorithm group that satisfies the most important property required by this scheme:

$$HFV = FHA(M) = FHA(C) = FHA(D(k_D, C)) = FHA(E(k_E, M))$$

Central projection is the projection of points on one line (curve, plane, surface) to another line (curve, plane, surface) from a central point. As shown in Fig 3, from the central point O, four points A, B, C, D on the X-axis are projected onto line l, resulting in another four points A'、B'、C'、D'.

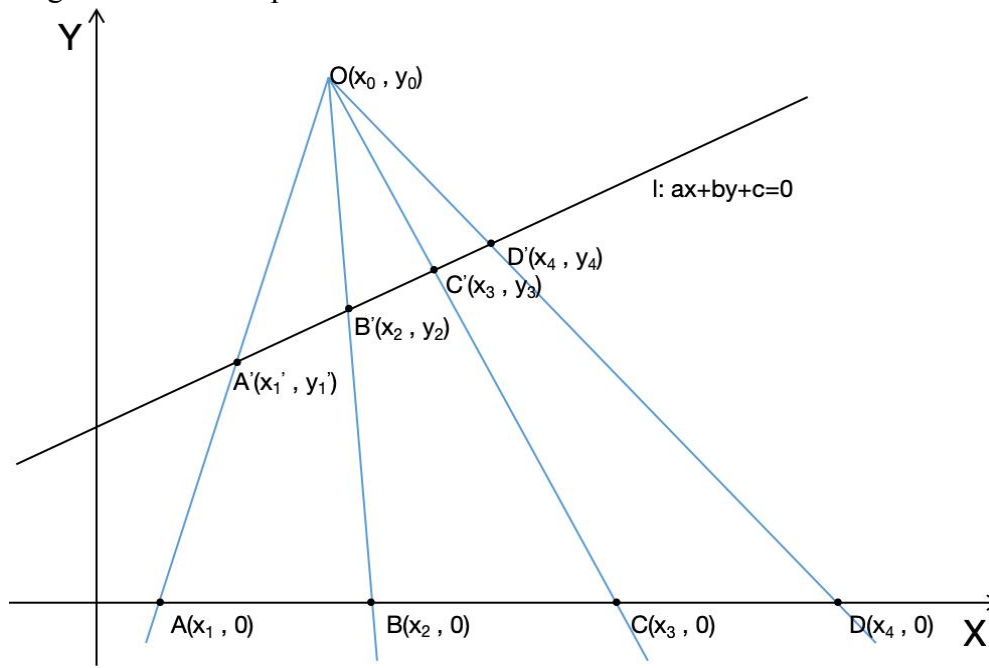

Fig 3 Central Projection and Cross-Ratio

The cross-ratio is defined as follows: for four points A, B, C, D on a straight line, the cross-ratio of these four points, denoted as ABCD, is defined by the following ratio:

$$\text{ABCD} = \frac{CA}{CB} \Big/ \frac{DA}{DB} \tag{7}$$

Under central projection, the cross-ratio of points A, B, C, D and their projected points A', B', C', D' remains invariant:

$$\text{ABCD} = \frac{CA}{CB} \Big/ \frac{DA}{DB} = A'B'C'D' = \frac{C'A'}{C'B'} \Big/ \frac{D'A'}{D'B'} \tag{8}$$

Constructing a data consistency verification algorithm group under ciphertext-only conditions using central projection and cross-ratio, the overall idea is as follows: use central projection to construct Feature Homomorphic Encryption and Decryption Algorithms E and D, and then use cross-ratio to construct the FHA. Since the cross-ratio remains invariant under central projection, this algorithm group can verify data consistency under ciphertext-only conditions. Additionally, because the cross-ratio is only related to the angles formed between lines and is independent of the positions of the eight points on the line or the distances between the points, it is impossible to deduce the original positions of the points from the cross-ratio, from a cryptographic perspective, it means that the plaintext cannot be deduced from the cross-ratio.

## 4.2. Constructing Feature Homomorphic Encryption and Decryption Algorithms through Central Projection

Using a single central projection to encrypt data essentially involves mapping points from the set {P | points on the x-axis with x-coordinates 0~N} to the set {Q | all points on the line ax+by+c=0} through a central projection centered at O. The coordinates of the mapped points are then serialized to form the ciphertext. Decryption is the reverse mapping process. The encryption algorithm E in the algorithm group is constructed with the following steps:

1) First, map the data to be encrypted to points on the X-axis:

Take the value of every n bits (8 bits, as an example, the same below) from the data to be encrypted as the X-axis coordinate, with the Y-axis coordinate being 0, as shown in Fig 3.

2) Second, select a central point O and a line l that meet the following conditions:

Choose a point $O(x_0, y_0)$ outside the X-axis as the central point of the projection, and arbitrarily draw a line $l: ax + by + c = 0$. This line should not pass through point O and should not be the X-axis (i.e., it should not be the line y = 0, or a and c should not both be 0).

3) Third, project the points on the X-axis onto line l:

Draw a line from the point on the X-axis, such as point A in Figure 1, through point O. The intersection point A' of this line with line l is the central projection of point A. The coordinates $(x_n', y_n')$ of point A' are determined by the following equation:

when $x_n \neq x_0$:

$$\begin{cases} x_n' = \dfrac{-cx_0 + cx_n + bx_n y_0}{ax_0 - ax_n + by_0} \\ y_n' = \dfrac{-ax_n y_0 - cy_0}{ax_0 - ax_n + by_0} \end{cases} \quad (9)$$

when $x_n = x_0$:

$$\begin{cases} x_n' = x_n \\ y_n' = -\dfrac{ax_n + c}{b} \end{cases} \quad (10)$$

when $(x_n - x_0)/y_0 = -b/a$:

$$\begin{cases} x_n' = x_n \\ y_n' = 0 \end{cases} \quad (11)$$

Since in cryptographic algorithms, all numbers must be integers, $x_n'$ and $y_n'$ should be reduced to their simplest fractional form, and then represented by listing the numerator and denominator side by side, that is, $x_n' \to$ (numerator:denominator). Here, we use $\underline{x_n}$ to represent the numerator and $\overline{x_n}$ to represent the denominator. Thus,

$x_n' \to (\underline{x_n'} : \overline{x_n'})$, $y_n' \to (\underline{y_n'} : \overline{y_n'})$.

In summary, the encryption key $k_E$ consists of the point $O(x_0, y_0)$ and the line $l: ax + by + c = 0$. In fact, the parameters $(x_0, y_0, a, b, c)$ together form the complete encryption key $k_E$.

The encryption algorithm $E(k_E, M)$ is:

**Step 1**, select the key $(x_0, y_0, a, b, c)$, where the key should meet the following conditions:

$$\begin{cases} ax_0 + by_0 + c \neq 0 & \text{(the point O is not on the projection line)} \\ y_0 \neq 0 & \text{(point O is not on the X-axis)} \\ |a| + |b| \neq 0 & \text{(a and c should not both be 0)} \end{cases} \quad (12)$$

**Step 2**, divide the original data to be encrypted into groups of 8 bits (i.e., 1 byte) each. Each group of data corresponds to $x_n$ as mentioned above.

**Step 3**, calculate the numerator and denominator of $x_n'$ and $y_n'$ for each $x_n$ data according to Equations (9), (10) or (11). Simultaneously, divide them by their greatest common divisor (GCD). The final result, after serialization, $(\underline{x_n'}:\overline{x_n'}:\underline{y_n'}:\overline{y_n'})$ is the encrypted data.

Constructing the decryption algorithm D involves re-projecting the points on line l back to the X-axis through point O. This is determined by the following equations:

when $x_n' \neq x_0$:

$$x_n = \frac{x_n' y_0 - x_0 y_n'}{y_0 - y_n'} = \frac{x_n' \overline{y_n'} y_0 - x_0 \overline{x_n'} y_n'}{x_n' \overline{y_n'} y_0 - \overline{x_n'} y_n'} \quad (13)$$

when $x_n' = x_0$ or $y_n' = 0$:

$$x_n = x_n' = \underline{x_n'} / \overline{x_n'} \quad (14)$$

When $y_n' = y_0$, thus $x_n = \infty$, since the plaintext should be a finite integer, this situation should not occur in practice.

Therefore, the decryption key $k_D$ is simpler than the encryption key $k_E$, and is given by $(x_0, y_0)$.

The decryption algorithm $D(k_D, C)$ is:

**Step 1**, use the received key $(x_0, y_0)$ and the ciphertext $(\underline{x_n'}:\overline{x_n'}:\underline{y_n'}:\overline{y_n'})$ to calculate $x_n$ using the Equations (13) or (14).

**Step 2**, repeat Step 1, and the calculated $(x_1, x_2, \cdots, x_n)$ will be the plaintext.

## 4.3. Constructing Feature Homomorphic Algorithm Group through Cross-Ratio

**Step 1**, calculate the cross-ratio for every four points.

For plaintext, calculate the cross-ratio for every 4 data using the following equation:

$$ABCD = \frac{x_3 - x_1}{x_3 - x_2} \bigg/ \frac{x_4 - x_1}{x_4 - x_2} = \frac{x_1 x_2 - x_1 x_4 - x_2 x_3 + x_3 x_4}{x_1 x_2 - x_1 x_3 - x_2 x_4 + x_3 x_4} \qquad (15)$$

For ciphertext, calculate the cross-ratio for every 16 data using the following equations:

$$A'B'C'D' = \frac{\sqrt{(x_3'-x_1')^2 + (y_3'-y_1')^2} \times \sqrt{(x_4'-x_2')^2 + (y_4'-y_2')^2}}{\sqrt{(x_3'-x_2')^2 + (y_3'-y_2')^2} \times \sqrt{(x_4'-x_1')^2 + (y_4'-y_1')^2}} \qquad (16)$$

$$A'B'C'D' = \frac{\sqrt{(x_3'\overline{x_1'}-x_1'\overline{x_3'})^2 \overline{y_1'}^2 \overline{y_3'}^2 + (y_3'\overline{y_1'}-y_1'\overline{y_3'})^2 \overline{x_1'}^2 \overline{x_3'}^2} \times \sqrt{(x_4'\overline{x_2'}-x_2'\overline{x_4'})^2 \overline{y_2'}^2 \overline{y_4'}^2 + (y_4'\overline{y_2'}-y_2'\overline{y_4'})^2 \overline{x_2'}^2 \overline{x_4'}^2}}{\sqrt{(x_3'\overline{x_2'}-x_2'\overline{x_3'})^2 \overline{y_2'}^2 \overline{y_3'}^2 + (y_3'\overline{y_2'}-y_2'\overline{y_3'})^2 \overline{x_2'}^2 \overline{x_3'}^2} \times \sqrt{(x_4'\overline{x_1'}-x_1'\overline{x_4'})^2 \overline{y_1'}^2 \overline{y_4'}^2 + (y_4'\overline{y_1'}-y_1'\overline{y_4'})^2 \overline{x_1'}^2 \overline{x_4'}^2}} \qquad (17)$$

Similarly, the cross-ratio used in cryptography cannot be in decimal form. To avoid non-integer results caused by square roots, both the numerator and denominator of the cross-ratio are squared. Then, after reducing the numerator and denominator to their simplest fractional form, they are represented side by side, i.e., ($\underline{ABCD}^2 : \overline{ABCD}^2$) and ($\underline{A'B'C'D'}^2 : \overline{A'B'C'D'}^2$), when the plaintext/ciphertext is less than 4/16, the cross-ratio is directly assigned a value of 1.

**Step 2**, using the HASH value of the cross-ratio as the HFV:

$$HASH\left(\underline{ABCD}^2 : \overline{ABCD}^2 : ...\right) \qquad (18)$$

$$HASH\left(\underline{A'B'C'D'}^2 : \overline{A'B'C'D'}^2 : ...\right) \qquad (19)$$

## 4.4. Algorithm Verification

Here, we verify the feasibility of the proposed algorithm through an example.

Using the above algorithm, select the plaintext as: "#Hello world!". The reason for adding a "#" at the beginning is that the first four characters of "Hello world!" (including the space) contain two equal characters "l", making the cross-ratio 1. The 5th to 8th characters contain two equal characters "o", making the cross-ratio infinite (which is generally directly assigned as 1 in practice). Therefore, adding the "#" makes the data more readable.

Select the encryption key $(x_0, y_0, a, b, c)$ as (5, 6, 2, -3, 4).

The ciphertext after encryption is:

85 13 74 13|257 38 111 19|239 35 206 35|383 56 165 28|383 56 165 28|787 115 678 115|13 2 17 3|281 41 242 41|787 115 678 115|404 59 348 59|383 56 165 28|355 52 153 26|241 37 210 37

Each group of 4 ciphertext data corresponds to 1 character of the plaintext. The encryption key is $(x_0, y_0)$ and is given as (5, 6). After decryption, the result is:

35 72 101 108 108 111 32 119 111 114 108 100 33

Corresponding exactly to the ASCII values of "#Hello world!".

The calculated plaintext cross-ratio is:

5645376 4481689 369664 755161 49 121 1 1

In the above data, every 2 numbers form a group corresponding to the cross-ratio of 4 characters in the plaintext. Since the last group of characters contains only the character "!", it is directly assigned the value "1 1".

The HASH value(hexadecimal, the same below) calculated from the cross-ratio data with spaces converted into a string is:

0x2998271ca969b65459c61bc9d4a2dd17cc9fd9b1298e17521433f072a9e90a28

The HASH value is the HFV calculated by this Feature Homomorphic Algorithm Group for "#Hello world!".

The calculated ciphertext cross-ratio is:

5645376 4481689 369664 755161 49 121 1 1

Every 2 numbers form a group corresponding to the cross-ratio of 16 numbers in the ciphertext. It can be seen that the cross-ratio of the plaintext and the ciphertext are equal. Therefore, the HFV of the ciphertext is calculated as:

0x2998271ca969b65459c61bc9d4a2dd17cc9fd9b1298e17521433f072a9e90a28

It can be seen that the HFV of the plaintext and ciphertext are equal. Technicians familiar with this field can verify this themselves. It should be noted that using different HASH algorithms may result in final HFV that are inconsistent with this scheme.

## 4.5. Security Analysis

Obviously, the above algorithm, as an exemplary and illustrative algorithm, does not meet cryptographic security requirements. However, analyzing the security of this algorithm is beneficial for subsequent security analysis after more complex modifications to the algorithm. The security of Feature Homomorphism must consider the security of the encryption and decryption algorithms, as well as the FHA from the perspective of a third-party attacker, and also from the perspective of the verifier. In this example, the attacker and the verifier are the same.

### 4.5.1. Known Ciphertext Attack

Easy to know, that this encryption method does not destroy the statistical properties of the plaintext. Given the ciphertext, the encryption algorithm can be immediately broken through statistical analysis.

In fact, very precisely, by first obtaining 2 sets of ciphertext, you can calculate the three parameters (a, b, c) based on the two-point line theorem. This means that the key has only two unknowns $(x_0, y_0)$. Including the plaintext, there are only 3 unknowns. Since 1 plaintext corresponds to 4 ciphertexts, you can set up 4 linear equations with 2 variables for these 3 unknowns, which allows you to solve for the key values $(x_0, y_0)$. Therefore, this method does not provide any security.

Additionally, under known ciphertext conditions, brute-force cracking is also not very difficult. Since the original mapped points are limited to N, the cipher parameters $(x_0, y_0, a, b, c)$ can be directly cracked using the theorem that 4 ciphertexts form a line through 2 points, which allows for the determination of the three parameters (a, b, c). Only the two parameters $(x_0, y_0)$ of point O need to be cracked. We can assume that the plaintext $x_n$ is one of the N points, then traverse the N points as $x_{n+1}$ to determine point O. By decrypting the other ciphertexts into plaintext, since the plaintext cannot be a non-integer, the parameters of point O can be quickly verified for correctness. Easy to know that at most $N^2$ attempts are needed to break this encryption algorithm. Since the value of N is limited, for example, in this paper N is less than 256, even using brute force, the security of the algorithm is insufficient.

#### 4.5.2. Known Plaintext Attack

Given the plaintext and the corresponding ciphertext, it can be seen from Equations (13) and (14) that the key $(x_0, y_0)$ has only 2 unknowns. With 1 set of plaintext and ciphertext, there are only 2 unknowns. By obtaining the ciphertext corresponding to 1 plaintext and setting up a system of linear equations with two variables, the key $(x_0, y_0)$ can be solved, indicating that the method is completely insecure.

#### 4.5.3. Collision Attack

Once the ciphertext is obtained, the attacker directly has access to the cross-ratio of each part of the ciphertext. Based on the method of central projection, it is easy to find data with the same cross-ratio to tamper with the ciphertext.

## 5. Feasible Scheme of Feature Homomorphic Algorithm Group

The Feature Homomorphic Algorithm Group provided above is merely a convenient example to illustrate the feasibility of Feature Homomorphism. It is intended to help readers understand and reproduce this scheme more easily and to design corresponding Feature Homomorphic Algorithm Group using the methods provided in this scheme. Clearly, it does not meet the cryptographic principles of confusion and diffusion, nor does it satisfy weak collision resistance. That is, from the obtained ciphertext, a third-party attacker can easily construct data with the same HFV to tamper with the data. Therefore, the cryptographic strength is extremely weak and does not meet practical requirements.

### 5.1. Approach to Make the Algorithm Feasible

The following provides ideas for enhancing the algorithm:

**First** approach is to use algorithms such as Cipher Feedback, Output Feedback, or Key Sequence during each round of encryption. By mixing the previous round's plaintext and key with the current round's plaintext to construct a new key for the current round, the statistical characteristics of the key and plaintext are confused and diffused throughout the entire ciphertext.

**Secondly**, in projective geometry, there are also cross-ratio theorems on conic sections and hyperbolic surfaces that can be incorporated into the algorithm to make it more nonlinear. Additionally, multiple and repeated projective transformations can be used, along with the introduction of modular arithmetic. By leveraging other mathematical challenges such as Large Integer Factorization Problem, Discrete Logarithm Problem, and Elliptic Curve Discrete Logarithm Problem, the algorithm can be modified to disrupt the linear relationship and one-to-one correspondence of cross-ratios or other characteristic values between the original plaintext and ciphertext.

This disruption of statistical relationships between data enhances collision resistance and provides stronger security and reliability.

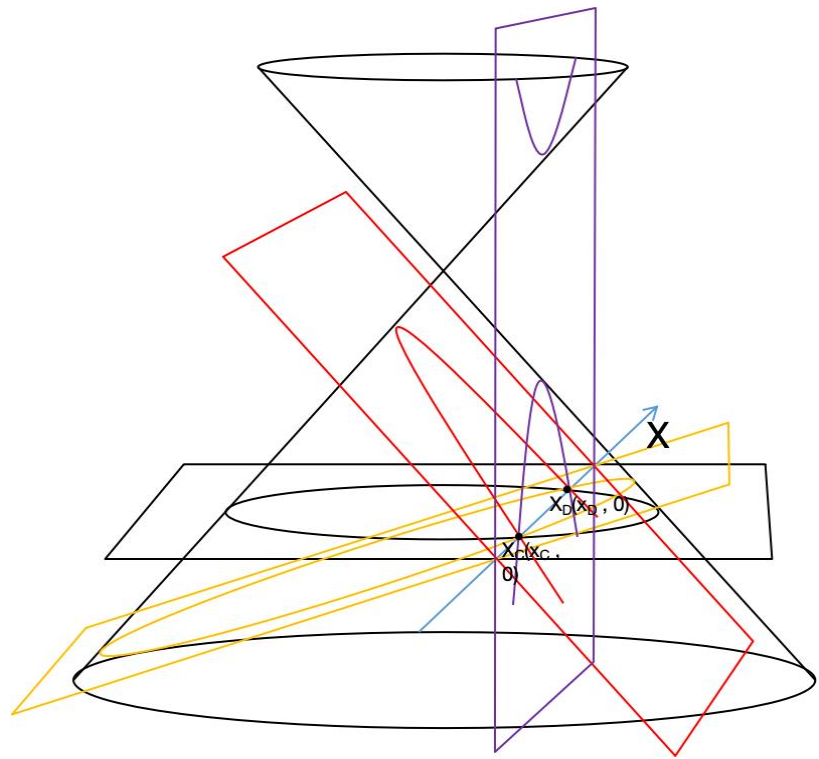

Fig 4 A diagram illustrating the projective geometric relationships between conic sections

As an invariant property of projective transformations, the cross-ratio theorem not only holds for points, lines, and planes but also for conic sections such as circles, ellipses, parabolas, and hyperbolas (Fig 4). There are also many different forms.

**Third**, change the algorithm process and design, such as adopting interactive methods involving data verifiers in plaintext or key generation, to enhance the security and reliability of the Feature Homomorphic Algorithm Group.

**Fourth**, other invariants can be found in various transformations (such as the Euler characteristic in topological transformations in geometry, eigenvalues, eigenvectors, and rank in certain matrix transformations in algebra, invariants in group, ring, and field structures in number theory, and even variants of information entropy in plaintext and ciphertext) to serve as the mathematical foundation for constructing Feature Homomorphic Algorithm Group.

Based on the above four points, this paper attempts to construct a more feasible Feature Homomorphic Algorithm Group.

## 5.2. Using Modular Arithmetic

Modular arithmetic is one of the most commonly used operations in cryptography. When applied to Feature Homomorphic Algorithm Group, it can effectively break their linear relationships.

Select any prime number p greater than N, then the modulo division operation satisfies the following equation (This equation does not always hold, but in this case, because the result can be confirmed to be an integer, it holds.):

$$x_n \% p = (\underline{x_n}/\overline{x_n})\%p = [\underline{x_n}\%p \times (\overline{x_n}^{-1})\%p]\%p \tag{20}$$

Let $|x_n'| = (\underline{x_n'}/\overline{x_n'})\%p$, $|y_n'| = (\underline{y_n'}/\overline{y_n'})\%p$, based on equation (20), although $|x_n'| \neq \underline{x_n'}/\overline{x_n'} = x_n'$, in this case (the calculation result must be an integer), the following equation holds:

$$x_n = (\frac{x_n'y_0 - x_0 y_n'}{y_0 - y_n'})\%p = (\frac{|x_n'|y_0 - x_0|y_n'|}{y_0 - |y_n'|})\%p \tag{21}$$

And the following equation:

$$x_n' = |x_n'| = (\underline{x_n'}/\overline{x_n'})\%p \tag{22}$$

Equations (21) and (22) correspond to equations (13) and (14), respectively.

So, taking the construction of the Feature Homomorphic Algorithm Group with central projection as an example, the encryption key with modulo operation is: $(x_0, y_0, a, b, c, p)$.

And the encryption steps are as follows: first, substitute the plaintext into equations (9) or (10) to obtain the numerator and denominator of the ciphertext, then perform modulo p division in equation (20), and finally calculate the resulting ciphertext as follows:

$$c_n = (|x_n'| : |y_n'|) \tag{23}$$

The only problem that needs to be solved is to ensure that the denominator calculated by the encryption algorithm has an inverse under modulo p. Since p is a prime number, this requires that the denominator cannot be a multiple of p. One method is to appropriately select the prime number p, or change p once it is found that the inverse does not exist.

The decryption key is (x0, y0, p). When decrypting, you only need to substitute the ciphertext into equations (21) or (22).

When the data owner/verifier calculates the HFV, only need to substitute the plaintext into equation (15)/ciphertext into equation (16) under modulo p, serialize it, and then compute the HASH value, which is the HFV of the plaintext/ciphertext introduced by the modulo operation.

## 5.3. Transform the Feature Homomorphic Algorithm Group Using the Principles of the ElGamal Algorithm

It is easy to know that the cryptographic strength of introducing modular arithmetic is still too low. However, after introducing modular arithmetic, the Feature Homomorphic Algorithm Group can be improved by some mathematical challenges, for example based on the principles of the ElGamal algorithm to further enhance cryptographic security.

### 5.3.1. Principles of the ElGamal Algorithm

ElGamal algorithm utilizes the discrete logarithm problem (DLP) to construct a highly secure public key algorithm. A brief introduction is as follows:

1) Select a large prime number p, ensuring that p-1 has a large prime factor. Then, randomly choose a generator g of the cyclic group $Z_p^*$;

2) Randomly select an integer x (requirement: $2 \leq x \leq p-2$), and calculate $y = g^x \pmod p$.

3) The private key is $(p, g, x)$, and the public key is $(p, g, y)$.

4) To encrypt plaintext m, randomly select an integer r ($2 \leq r \leq p-2$ and coprime with p-1), calculate $c1 = g^r \pmod p$, $c1 = m \times y^r \pmod p$, and obtain the ciphertext $C = (c1, c2)$;

5) Decrypt the plaintext $m = c2 \times c1^{-x} \pmod p$:

$$c2 \times c1^{-x} \% p \equiv m \times y^r \times g^{-xr} \equiv m \times g^{xr} \times g^{-xr} \equiv m \qquad (24)$$

### 5.3.2. Transform the Encryption Algorithm

**Step 1**, According to the requirements of section 5.3.1, select p and generator g, randomly choose x, let $y = g^x \pmod p$, the encryption key is: $(x_0, y_0, a, b, c, p, g, y)$.

**Step 2**, According to the specific situation, substitute the plaintext m into equations (9), (10) or (11), and respectively calculate the numerators and denominators of $x_n{'}$ and $y_n{'}$, then perform modulo p division to obtain $|x_n'|$, $|y_n'|$.

**Step 3**, randomly choose r, x and y are respectively modulo multiplication by $y^r$ to obtain the ciphertext:

$$c_n = (|x_n'|y^r \% p : |y_n'|y^r \% p) \qquad (25)$$

If, as in section 4.2, the numerator and denominator are respectively multiplied by $y^r$ to form the ciphertext, then from equations (13) and (14), the encryption factor $y^r$ can be directly canceled out by modulo division. The strength of the cipher would then rely solely on $x_0$ and $y_0$, significantly reducing its security.

### 5.3.3. Transform the Decryption Algorithm

The decryption key is: $(x_0, y_0, a, b, c, p, g, x)$.

There are two ways to decrypt. The first is ciphertext modulo p multiply $c1^{-x}$, then substitute into equation (21) to decrypt the plaintext.

The second method, observation equation (21), only requires multiplying $y_0$ in the denominator by $y^r$, $y^r$ in the equation will be eliminated. and decrypt the plaintext:

$$x_n = (\frac{|x_n'|y^r y_0 - x_0|y_n'|y^r}{y_0 y^r - |y_n'|y^r}) \% p = (\frac{|x_n'|y_0 - x_0|y_n'|}{y_0 - |y_n'|}) \% p \qquad (26)$$

### 5.3.4. Transform FHA

The plaintext is directly substituted into equation (15) (take the square, using modulo p division) for cross-ratio calculation. After serializing the obtained cross-ratio, a hash operation is performed to get the plaintext HFV.

To obtain the HFV from the ciphertext, during encryption, a single key is selected for every group of 4 data points. The ciphertext is directly substituted into the cross-ratio equation (16) (take the square, using modulo p division), and after serializing the obtained cross-ratio, a hash operation is performed to get the ciphertext HFV.

When the plaintext/ciphertext is less than 4/8, the cross-ratio is directly assigned a value of 1.

### 5.3.5. Algorithm Verification

The plaintext remains: "#Hello world!". Select p=167, g=83, x=16, r=13. It should be noted that the method of changing r each time is not adopted here, make the comparison of data before and after more obvious.

The encryption key $(x_0, y_0, a, b, c, p, g, y)$ is (5, 6, 2, -3, 4, 167, 83, 58).

The ciphertext after encryption is:

41 130 | 156 151 | 125 19 | 130 78 | 130 78 | 161 43 | 83 158 | 153 149 | 161 43 | 114 123 | 130 78 | 60 87 | 154 94

Each group of two corresponds to one character in plaintext.

The decryption key $(x_0, y_0, a, b, c, p, g, x)$ is (5, 6, 2, -3, 4, 167, 83, 16). After decryption, the result is:

35 72 101 108 108 111 32 119 111 114 108 100 33

The data correspond exactly to the ASCII code values of "#Hello world!"

The plaintext ratio is:

99 147 126 1

Each number corresponds to the cross-ratio of 4 characters in the plaintext. After converting the cross-ratio data with spaces into a string, then calculate the HASH value (plaintext HFV) is:

0x72be28af954a135b7d6f7317f6da941dd1b56cba2e2b45f0c20259f4ed1fdfd6

The calculated ciphertext cross-ratio is:

99 147 126 1

Each number corresponds to the cross-ratio of 8 datas in the ciphertext. It can be seen that the cross-ratio of the plaintext and the ciphertext are equal. Therefore, the HFV of the ciphertext is:

0x72be28af954a135b7d6f7317f6da941dd1b56cba2e2b45f0c20259f4ed1fdfd6

It can be seen that the plaintext and ciphertext have equal HFV.

### 5.3.6. Security Analysis

Due to the requirements of privacy computing, the transformed homomorphic algorithm cannot be used as a public key cryptosystem. However, for the verifier and Attacker, when the key is well protected, its security is equivalent to the cryptographic strength of the ElGamal algorithm when the public key is unknown. The paper[37] analyzes the security of the ElGamal algorithm, which readers can refer to.

Due to the characteristics of cross-ratio, it is very easy for data groups and HFV within the data groups to collide. In fact, once the 4 data of data group is determined, the cross-ratio data is determined by p:

$$(B \times C^{-1})\%p = (B \times C^{p-2})\%p \tag{27}$$

For example, let's compare the 5.2 algorithm with the 5.3 algorithm using the plaintext: "#Hello world!". Similarly, the encryption key $(x_0, y_0, a, b, c, p, g, y)$ is (8, 12, 25, 78, 34, 167), which means the same p is selected.

The ciphertext after encryption is:

106 154 | 24 9 | 22 91 | 31 56 | 31 56 | 146 17 | 20 6 | 36 18 | 146 17 | 165 73 | 31 56 | 124 84 | 65 165

It is easy to see that the ciphertext changes significantly, but the calculated HFV value is:

0x72be28af954a135b7d6f7317f6da941dd1b56cba2e2b45f0c20259f4ed1fdfd6

If we change p to 100043, the HFV will be:

0x152b6767444befa829203bc69fdf9e01aa8aadd969ad704811e53f818213282f

The numerical values are completely different, verifying that once the data is determined in groups of 4, the cross-ratio data is determined by p.

Therefore, after obtaining the HFV, it is easy to perform a brute-force attack. Taking the 8-bit data in this article as an example, to fully obtain the HFV of the data, it only requires:

$$P_{256}^4 = \frac{256!}{252!} = 4195023360$$

Approximately 4.2 billion calculations are required to obtain a HFV table for a certain p-value. If the data is known to be in a specific language, such as English, even fewer calculations are needed. Additionally, due to the cross-ratio characteristics, the plaintext adding, subtracting, multiplying, or dividing by a certain number will result in the same cross-ratio, which makes it less resistant to collisions but relatively increases the difficulty of cracking.

## 5.4. Feature Homomorphic Algorithm Group With Random Array

Considering the possibility of brute-force attacks on cryptographic algorithms, we can also introduce random noise to disrupt the statistical properties of the data, thereby enhancing the security of Feature Homomorphic Algorithm Group.

### 5.4.1. Transform the Encryption Algorithm

As an example, the following introduces a scheme that multiplies the original data by a random factor to add noise.

1) First, select a random number $r_v = (r_{v1}, r_{v2}, \cdots)$. The encryption key is: $(x_0, y_0, a, b, c, p, g, y, r_v)$.

2) When encrypting, select one (or more) from a set of four data points and multiply it (or them) with one (or more) elements from the random array $r_v$. Substitute the result of the multiplication into the algorithm in Section 4.2. The modified equations is as follows:

when $x_n \neq x_0$:

$$\begin{cases} x_n' = \dfrac{-cx_0 + cr_{vn}x_n + br_{vn}x_n y_0}{ax_0 - ar_{vn}x_n + by_0}\%p \\ y_n' = -\dfrac{ar_{vn}x_n y_0 + cy_0}{ax_0 - ar_{vn}x_n + by_0}\%p \end{cases} \quad (28)$$

when $x_n = x_0$:

$$\begin{cases} x_n' = (r_{vn}x_n)\%p \\ y_n' = (-\dfrac{ar_{vn}x_n + c}{b})\%p \end{cases} \quad (29)$$

when $(x_n - x_0)/y_0 = -b/a$:

$$\begin{cases} x_n' = (r_{vn}x_n)\%p \\ y_n' = 0 \end{cases} \quad (30)$$

If $c_n^r$ represents the ciphertext using the equations (28), (29) or (30) (taking the first of the 4 data points multiplied by a random factor as an example), the transformed ciphertext is:

$$C = (c_1^r, c_2, c_3, c_4, c_5^r, c_6, c_7, c_8, \cdots) \quad (31)$$

### 5.4.2. Transform the Decryption Algorithm

When decrypting, the corresponding factor $(r_v^{p-2})\%p$ needs to be added.

Therefore, the key is: $(x_0, y_0, a, b, c, p, g, x, (r_v^{p-2})\%p)$.

To decrypt the ciphertext, substitute it into Equations (21) or (22) in Section 5.2, and perform a modulo multiplication of the ciphertext $c_n^r$ by $(r_v^{p-2})\%p$ modulo p.

### 5.4.3. Transform FHA

When calculating the plaintext cross-ratio, the factor $r_{vn}$ also needs to be included, as follows:

$$ABCD = \dfrac{r_{v1}x_1 x_2 - r_{v1}x_1 x_4 - x_2 x_3 + x_3 x_4}{r_{v1}x_1 x_2 - r_{v1}x_1 x_3 - x_2 x_4 + x_3 x_4}\%p \quad (32)$$

To calculate the ciphertext cross-ratio, simply substitute the ciphertext directly into the cross-ratio equations (16). Let $\|m_n - m_{n-1}\|$ represent $\sqrt{(x_n' - x_{n-1}')^2 + (y_n' - y_{n-1}')^2}$, then the ciphertext cross-ratio is as follows:

$$A'B'C'D' = \dfrac{\|m_3 - m_1^r\| \times \|m_4 - m_2\|}{\|m_3 - m_2\| \times \|m_4 - m_1^r\|}\%p \quad (33)$$

According to the cross-ratio theorem, equations (32) and (33) are equal. It is particularly important to note that when using equations (32) and (33) to calculate the cross-ratio, you need to take the square.

### 5.4.4. Algorithm Verification

The plaintext remains: "#Hello world!". Select p=167, g=83, x=16, r=13, $r_{vn}$ =39.

The encryption key $(x_0, y_0, a, b, c, p, g, y, r_{vn})$ is (5, 6, 2, -3, 4, 167, 83, 58, 39).

The ciphertext after encryption is:

163 100 | 156 151 | 125 19 | 130 78 | 49 24 | 161 43 | 83 158 | 153 149 | 25 8 | 114 123 | 130 78 | 60 87 | 2 104

Each group of 2 data corresponds to 1 character in plaintext.

The decryption key $(x_0, y_0, a, b, c, p, g, x, (r_v^{p-2})\%p)$ is (5, 6, 2, -3, 4, 167, 83, 16, 30). After decryption, the result is:

35 72 101 108 108 111 32 119 111 114 108 100 33

The data correspond exactly to the ASCII code values of "#Hello world!"

However, if the factor $r_v$ is ignored, the decryption result is:

29 72 101 108 37 111 32 119 154 114 108 100 118

The cross-ratio of plaintext is:

19 124 126 1

The HFV of the plaintext is:

0xf71812172fa29b18a4bde91adae53b73f9510cc0b2fdb3aacc2032f67920c981

The calculated ciphertext cross-ratio is:

19 124 126 1

The HFV of the ciphertext is:

0xf71812172fa29b18a4bde91adae53b73f9510cc0b2fdb3aacc2032f67920c981

It can be seen that the plaintext and ciphertext have equal HFV. However, HFV is significantly different from the original data.

Additionally, an interesting situation can be observed here. Although the first data in each group of four was multiplied by 39, the cross-ratio of the third group still resulted in a collision. This type of collision is clearly impossible without the introduction of modular arithmetic, and it is evident that this is largely related to our choice of a relatively small prime number p. If we choose a slightly larger prime number, p=100043, and recalculate the cross-ratio, the original data cross-ratio would be:

34459 3457 97563 1

The cross-ratio of data with a random factor is:

34552 97801 56962 1

The collision has disappeared.

## 6. The Universality of Feature Homomorphism Algorithms

When directly using algorithms such as ElGamal for encryption, it is actually a special case of central projection, that is, projecting points on the X-axis from any point outside the X-axis onto the X-axis. The points of the original data will still project onto their original points, and at this case, the cross-ratio theorem obviously still holds. Here, taking the ElGamal algorithm as an example, provid scheme for constructing FHA.

## 6.1. Constructing FHA Using Common Factor Subtraction

### 6.1.1. Transform FHA

Since the X-axis onto the X-axis projection is used, the cross-ratio of the plaintext and ciphertext both satisfy equations (15). Let the random factor set selected for the ciphertext be $R = (r_1, r_2, \cdots)$, then the ciphertext is directly substituted into equations (15) to obtain:

$$A'B'C'D' = \frac{y^{r_1}x_1 y^{r_2}x_2 - y^{r_1}x_1 y^{r_4}x_4 - y^{r_2}x_2 y^{r_3}x_3 + y^{r_3}x_3 y^{r_4}x_4}{y^{r_1}x_1 y^{r_2}x_2 - y^{r_1}x_1 y^{r_3}x_3 - y^{r_2}x_2 y^{r_4}x_4 + y^{r_3}x_3 y^{r_4}x_4} \%p \qquad (34)$$

Obviously, it is different from the plaintext cross-ratio directly calculated by equations (15). Therefore, if we consider randomly selecting a common factor $k^f$ (Note that an appropriate $k_f$ should be selected to satisfy $2 \leq r \leq p-2$ and coprime with p-1) and constructing the key for HFV as follows:

$$\begin{cases} k_{f1} = k_f - r_1 - r_2 \\ k_{f2} = k_f - r_3 - r_4 \\ k_{f2} = k_f - r_1 - r_4 \\ k_{f4} = k_f - r_1 - r_3 \end{cases} \qquad (35)$$

When sending the ciphertext, the HFV key $(y^{k_{f1}}, y^{k_{f2}}, y^{k_{f3}}, y^{k_{f4}})$ is also sent to the verifier. The verifier calculates the ciphertext according to the following equation:

$$\begin{aligned} A'B'C'D' &= \frac{x_1'x_2'y^{-k_{f1}} - x_1'x_4'y^{-k_{f3}} - x_2'x_3'y^{-k_{f1}-k_{f2}+k_{f3}} + x_3'x_4'y^{-k_{f2}}}{x_1'x_2'y^{-k_{f1}} - x_1'x_3'y^{-k_{f4}} - x_2'x_4'y^{-k_{f1}-k_{f2}+k_{f4}} + x_3'x_4'y^{-k_{f2}}} \\ &= \frac{y^{r_1}x_1 y^{r_2}x_2 y^{-k_{f1}} - y^{r_1}x_1 y^{r_4}x_4 y^{-k_{f3}} - y^{r_2}x_2 y^{r_3}x_3 y^{-k_{f1}-k_{f2}+k_{f3}} + y^{r_3}x_3 y^{r_4}x_4 y^{-k_{f2}}}{y^{r_1}x_1 y^{r_2}x_2 y^{-k_{f1}} - y^{r_1}x_1 y^{r_3}x_3 y^{-k_{f4}} - y^{r_2}x_2 y^{r_4}x_4 y^{-k_{f1}-k_{f2}+k_{f4}} + y^{r_3}x_3 y^{r_4}x_4 y^{-k_{f2}}} \\ &= \frac{y^{k_f}x_1 x_2 - y^{k_f}x_1 x_4 - y^{k_f}x_2 x_3 + y^{k_f}x_3 x_4}{y^{k_f}x_1 x_2 - y^{k_f}x_1 x_3 - y^{k_f}x_2 x_4 + y^{k_f}x_3 x_4} \\ &= \frac{x_1 x_2 - x_1 x_4 - x_2 x_3 + x_3 x_4}{x_1 x_2 - x_1 x_3 - x_2 x_4 + x_3 x_4} \%p \end{aligned} \qquad (36)$$

The cross-ratio that matches the plaintext can be obtained from the ciphertext.

### 6.1.2. Algorithm Verification

The plaintext remains: "#Hello world!". Select p=100043, g=83, x=16, r=[11, 31, 23, 53, 42, 18, 26, 34, 9, 57, 73, 82, 45], $k_f$ =[157, 593, 348].

The encryption key $(p, g, y)$ is (100043, 83, 11046), r is [11, 31, 23, 53, 42, 18, 26, 34, 9, 57, 73, 82, 45].

The ciphertext after encryption is:

5308 53413 65797 65065 11191 96387 41219 39237 54656 48716 74865 92388 20022

Each 1 data corresponds to 1 character in plaintext.

The decryption key $(p, g, x)$ is (100043, 83, 16). c1 is (43944, 92349, 45859, 65239, 40204, 63214, 9604, 61747, 71920, 38601, 3380, 85153, 43922). After decryption, the result is:

35 72 101 108 108 111 32 119 111 114 108 100 33

The data correspond exactly to the ASCII code values of "#Hello world!"

The $(y^{k_{f1}}, y^{k_{f2}}, y^{k_{f3}}, y^{k_{f4}})$ is (18128, 97714, 64995, 41197), (60780, 60780, 14578, 60160) and (74958, 29962, 84448, 55033). The cross-ratio of plaintext is:

48345 91985 81854 1

The HFV of the plaintext is:

0xed1d385bee5884b7676b22e2d9e4637925b9a6c9309d02e48065eba92d267c6c

The calculated ciphertext cross-ratio is:

48345 91985 81854 1

The HFV of the ciphertext is:

0xed1d385bee5884b7676b22e2d9e4637925b9a6c9309d02e48065eba92d267c6c

It can be seen that the plaintext and ciphertext have equal HFV.

The problem with this scheme is that it is prone to brute force attacks. Furthermore, as mentioned in Section 5.4, a random number $r_v$ can be added. For example, the HFV key can be modified to $(r_v y^{k_{f1}}, y^{k_{f2}}, y^{k_{f3}}, y^{k_{f4}})$, and the HFV of the plaintext feature is calculated according to the following equation:

$$ABCD = \frac{r_v^{-1} x_1 x_2 - x_1 x_4 - r_v^{-1} x_2 x_3 + x_3 x_4}{r_v^{-1} x_1 x_2 - x_1 x_3 - r_v^{-1} x_2 x_4 + x_3 x_4} \% p \qquad (37)$$

The equations for the HFV of ciphertext is as follows:

$$A'B'C'D' = \frac{x_1' x_2' r_v^{-1} y^{-k_{f1}} - x_1' x_4' y^{-k_{f3}} - x_2' x_3' r_v^{-1} y^{-k_{f1}-k_{f2}+k_{f3}} + x_3' x_4' y^{-k_{f2}}}{x_1' x_2' r_v^{-1} y^{-k_{f1}} - x_1' x_3' y^{-k_{f4}} - x_2' x_4' r_v^{-1} y^{-k_{f1}-k_{f2}+k_{f4}} + x_3' x_4' y^{-k_{f2}}} \% p \qquad (38)$$

This not only further damages the structure of the cross-ratio but also reduces the possibility of data leakage caused by providing 4 HFV key.

## 6.2. Feature Homomorphic Algorithm Group With Verifier's Random Array

Here we introduce a method to construct a FHA by incorporating a random array provided by the verifier, and altering the verification process.

### 6.2.1. Transform the Algorithm Group

We will construct the Feature Homomorphic Algorithm Group according to the following approach:

1) First, the data verifier randomly selects a set of random numbers r and sends them to the data owner through a secure method (encryption, trusted channel, or key exchange, etc.).

2) Use the original ElGamal algorithm to compute the ciphertext, as follow:

$$c_n = (m \times y^r) \% p \qquad (39)$$

Ciphertext is:

$$C = (c_1, c_2, c_3, c_4, \cdots) \qquad (40)$$

Decryption also uses the original ElGamal algorithm.

3) When calculating the plaintext cross-ratio, random factor $r_{vn}$ needs to be added, as follows:

$$ABCD = \frac{r_{v1}x_1x_2 - r_{v1}x_1x_4 - x_2x_3 + x_3x_4}{r_{v1}x_1x_2 - r_{v1}x_1x_3 - x_2x_4 + x_3x_4} \% p \tag{41}$$

Since no factor was added during encryption, when calculating the ciphertext cross-ratio, the verifier needs to add the random factor $r_{vn}$, as follows:

$$A'B'C'D' = \frac{r_v^{-1}x_1'x_2'y^{-k_{f1}} - r_v^{-1}x_1'x_4'y^{-k_{f3}} - x_2'x_3'y^{-k_{f1}-k_{f2}+k_{f3}} + x_3'x_4'y^{-k_{f2}}}{r_v^{-1}x_1'x_2'y^{-k_{f1}} - r_v^{-1}x_1'x_3'y^{-k_{f4}} - x_2'x_4'y^{-k_{f1}-k_{f2}+k_{f4}} + x_3'x_4'y^{-k_{f2}}} \% p$$

$$= \frac{r_v^{-1}x_1x_2 - r_v^{-1}x_1x_4 - x_2x_3 + x_3x_4}{r_v^{-1}x_1x_2 - r_v^{-1}x_1x_3 - x_2x_4 + x_3x_4} \% p \tag{42}$$

First, the plaintext and ciphertext have equal cross-ratio when calculated this way. Secondly, with the random array kept confidential, a third party cannot compute this cross-ratio, thus preventing the use of cross-ratio brute force attacks to crack the encryption algorithm. This greatly enhances the security of the algorithm.

### 6.2.2. Data Validation Process

The data consistency verification scheme for the FHA of mixed verifier's random array is shown in Fig 5:

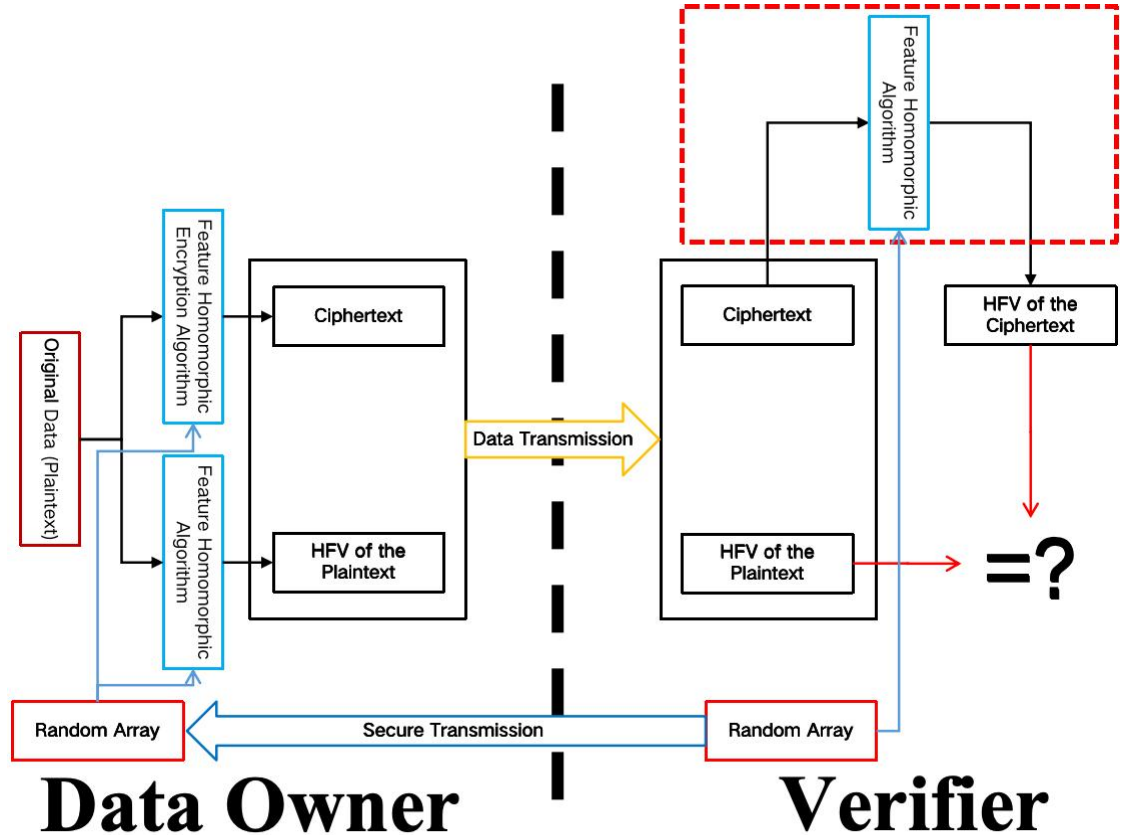

Fig 5 Schematic diagram for data consistency verification process based on the FHA of mixed verifier's random arrays

Compared with Fig 2, the differences are: first, a random array generated by the verifier is added, and the confidentiality of the random array needs to be ensured;

second, when generating plaintext and ciphertext HFV, the random array needs to be added; third, the encryption and decryption algorithm still uses the original ElGamal algorithm. Therefore, security, convenience, and collision resistance are all improved.

## 6.3. Constructing FHA Using Equalize Sums

If we divide the plaintext data into several parts, and each part is divided into 4 groups according to an agreed method, with each group containing a certain number of data points, then when selecting the encryption factors for these 4 groups, it is required that the sum of the factors of the 4 groups in each part be equal. In this way, the product of these 4 groups of data can be used as new data of four new points to construct the cross-ratio ABCD, which can be used to construct a Feature Homomorphic Algorithm Group.

### 6.3.1. Transform the Algorithm Group

Let $r_S$ be the sum, and the encryption factors of each data of points A, B, C, and D are $r_{A1}$, $r_{A2}$, $r_{A3}$, …, $r_{B1}$, $r_{B2}$, $r_{B3}$, …, $r_{C1}$, $r_{C2}$, $r_{C3}$, …, $r_{D1}$, $r_{D2}$, $r_{V3}$, … respectively, satisfying the following relationship:

$$\begin{cases} r_S = r_{A1} + r_{A2} + r_{A3} + \cdots \\ r_S = r_{B1} + r_{B2} + r_{B3} + \cdots \\ r_S = r_{C1} + r_{C2} + r_{C3} + \cdots \\ r_S = r_{D1} + r_{D2} + r_{D3} + \cdots \end{cases} \tag{43}$$

Note that the above equations is just an example of the encryption factors for ABCD, and it does not mean that there must be more than 3 or that they must be equal in number. And, it only needs to meet the basic requirements of r (include $r_S$), and r should be selected as randomly as possible. Additionally, it is necessary to fill in the data according to the agreed data grouping method.

Use the ElGamal algorithm to encrypt and decrypt data. The plaintext of the 4 points ABCD in the new algorithm is:

$$\begin{cases} x_A = x_{A1} \times x_{A2} \times x_{A3} \times \cdots \\ x_B = x_{B1} \times x_{B2} \times x_{B3} \times \cdots \\ x_C = x_{C1} \times x_{C2} \times x_{C3} \times \cdots \\ x_D = x_{D1} \times x_{D2} \times x_{D3} \times \cdots \end{cases} \tag{44}$$

The plaintext ratio is calculated using the following equation:

$$ABCD = \frac{x_A x_B - x_A x_D - x_B x_C + x_C x_D}{x_A x_B - x_A x_C - x_B x_D + x_C x_D} \% p \tag{45}$$

The ciphertext ratio is calculated using the following equation:
Let:

$$\begin{cases} x_A' = x_{A1}' x_{A2}' x_{A3}' \cdots = x_{A1} x_{A2} x_{A3} \cdots \times y^{r_{A1}} y^{r_{A2}} y^{r_{A3}} \cdots = x_{A1} x_{A2} x_{A3} \cdots \times y^{r_S} \\ x_B' = x_{B1}' x_{B2}' x_{B3}' \cdots = x_{B1} x_{B2} x_{B3} \cdots \times y^{r_{B1}} y^{r_{B2}} y^{r_{B3}} \cdots = x_{B1} x_{B2} x_{B3} \cdots \times y^{r_S} \\ x_C' = x_{C1}' x_{C2}' x_{C3}' \cdots = x_{C1} x_{C2} x_{C3} \cdots \times y^{r_{C1}} y^{r_{C2}} y^{r_{C3}} \cdots = x_{C1} x_{C2} x_{C3} \cdots \times y^{r_S} \\ x_D' = x_{D1}' x_{D2}' x_{D3}' \cdots = x_{D1} x_{D2} x_{D3} \cdots \times y^{r_{D1}} y^{r_{D2}} y^{r_{D3}} \cdots = x_{D1} x_{D2} x_{D3} \cdots \times y^{r_S} \end{cases}$$

then:

$$A'B'C'D' = \frac{x_A'x_B' - x_A'x_D' - x_B'x_C' + x_C'x_D'}{x_A'x_B' - x_A'x_C' - x_B'x_D' + x_C'x_D'} \% p$$
$$= \frac{x_A x_B y^{2S} - x_A x_D y^{2S} - x_B x_C y^{2S} + x_C x_D y^{2S}}{x_A x_B y^{2S} - x_A x_C y^{2S} - x_B x_D y^{2S} + x_C x_D y^{2S}} \% p \quad (46)$$
$$= \frac{x_A x_B - x_A x_D - x_B x_C + x_C x_D}{x_A x_B - x_A x_C - x_B x_D + x_C x_D} \% p$$

### 6.3.2. Algorithm Verification

As an example, here, take one data point every 4, and group every 4 data points into one set, and the sum of the encryption factors for each set of 4 data points is equal.

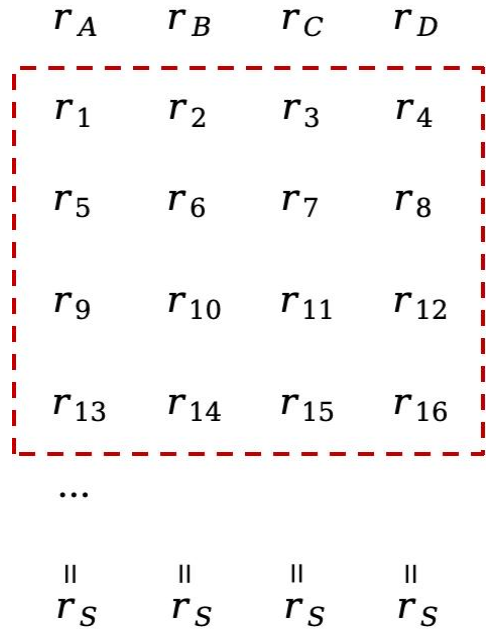

Fig 6 4×4 data block and random number r relationship diagram

The plaintext padding to 16 datas: "#Hello world!{|}". Select p=100043, g=83, x=16, r=r1=[11, 31, 23, 53, 23,15,17,10, 36, 21, 25, 9, 30, 33, 35, 28], $r_S$ =100.

The encryption key $(p, g, y)$ is (100043, 83, 11046), r is [11, 31, 23, 53, 23,15,17,10, 36, 21, 25, 9, 30, 33, 35, 28]).

The ciphertext after encryption is:

35380 83 16766 38951 17928 52226 88421 35191 79436 80961 66944 72379 35445 57344 74581 46267

Each 1 data corresponds to 1 character in plaintext.

The decryption key $(p, g, x)$ is (100043, 83, 16). After decryption, the result is:
35 72 101 108 108 111 32 119 111 114 108 100 33 123 124 125

The data correspond exactly to the ASCII code values of "#Hello world!{|}"

The cross-ratio of plaintext is:
39055

The HFV of the plaintext is:
0x1ad9f165063877b614906c72b1b0628b2b4f1ccc8ff45097a60e051fd80d44cb

The calculated ciphertext cross-ratio is:
39055
The HFV of the ciphertext is:
0x1ad9f165063877b614906c72b1b0628b2b4f1ccc8ff45097a60e051fd80d44cb
It can be seen that the plaintext and ciphertext have equal HFV.

### 6.4. Verification of Multiplicative Homomorphism and Feature Homomorphism of the Algorithm

We know that the ElGamal algorithm satisfies the multiplicative homomorphism. For now, regardless of whether it makes sense, let's modulo p multiply the ciphertexts from section 6.1, items 1 to 4, with the ciphertexts from items 5 to 8 (i.e., $x_1 \cdot x_5$, $x_2 \cdot x_6$, $x_3 \cdot x_7$, $x_4 \cdot x_8$), is:
76329, 6008, 20856, 58131

c1 needs to be converted into $c1(0\sim3) \times c1(4\sim7)$, is:
65239, 40550, 40550, 81138

Decrypt to:
3780 7992 3232 12852

These are exactly equal to the product of the plaintexts directly multiplied. Now we calculate the HFV of these two sets of data.

To calculate the plaintext HFV, you need to add $y^{k_{f(0\sim3)}+k_{f(4\sim7)}}$: [46281, 4225, 89900, 46281]. The cross-ratio is calculated as:
83094

The HFV of the plaintext is:
0xfd3512dcfa51ca394c38d664ecbfa42123f80e75af64e81c1e059439f4241837

The calculated ciphertext cross-ratio is:
83094

The HFV of the ciphertext is:
0xfd3512dcfa51ca394c38d664ecbfa42123f80e75af64e81c1e059439f4241837

The plaintext and ciphertext HFV are still equal, which means that even after homomorphic computation, the HFV of the ciphertext and plaintext remain equal, and the FHA is still effective.

So, we can assert that as long as the encryption algorithm is based on arithmetic encryption, it is possible to construct a FHA based on cross-ratio theorem to calculate the same HFV for both plaintext and ciphertext. In other words, the FHA is universal. And algorithm that satisfies both Feature Homomorphism and Computational Homomorphism exists. Additionally, any fully homomorphic algorithm can construct a FHA through the cross-ratio theorem.

### 7. Conclusion

This paper proposes a new homomorphism, Feature Homomorphism, which can calculate the same eigenvalues from plaintext-ciphertext pairs. It also presents a specific scheme for implementing the Feature Homomorphic Algorithm Group using cross-ratio and proposes various ideas to improve security and practicality. This scheme can be applied at least to data consistency and integrity verification in scenarios where "Data Availability versus Visibility", indexing of encrypted data, and for implementing zero-knowledge proofs.

This paper mainly provides ideas and methods for data consistency and integrity verification, searchable encryption, and zero-knowledge proofs in privacy computing. Therefore, it does not rigorously analyze security and efficiency, nor does it design the algorithms for engineering usability. It aims to inspire subsequent researchers to gain insights from it, and through further improvements, optimizations, or transformations, or by mixing various design ideas, design usable algorithms to meet more data application scenarios.